\documentclass{aa} % ``standard'' version
\usepackage{graphicx}
\usepackage{epsfig}
\usepackage{natbib}
\bibpunct{(}{)}{;}{a}{}{,}
\begin{document}

%   \thesaurus{06     % A&A Section 6: Form. struct. and evolut. of stars
%              (03.13.2;  % Methods: data analysis,
%               02.12.2)} % Line: identification,
%
%   \thesaurus{15     % astronomical instrumentation, methods and techniques
%              (15.13.2;  % Methods: data analysis,
%               02.12.2)} % Line: identification,
%

\title{Monte Carlo Study of Detector Concepts for the MAX Laue Lens
Gamma-Ray Telescope}

%   \subtitle{I. Overviewing the $\kappa$-mechanism}

\author{G. Weidenspointner\inst{1}
        \and
	C.B. Wunderer\inst{2}
        \and
        N. Barri\`ere\inst{1}
        \and
        A. Zoglauer\inst{2}
        \and
        P. von Ballmoos\inst{1}
}

\offprints{G. Weidenspointner, e-mail Georg.Weidenspointner@cesr.fr}

\institute{Centre d'Etude Spatiale des Rayonnements, 9 Avenue Colonel
           Roche, 31028 Toulouse Cedex 4, France
           \and
           Space Science Laboratory, UC Berkeley, Berkeley, USA
}

\date{Received ; accepted }

% because title is too long

   \titlerunning{Study of MAX Detector Concepts}
   \authorrunning{G. Weidenspointner et al.}

%%%   abstract   %%%

\abstract{
MAX is a proposed Laue lens gamma-ray telescope taking advantage of
Bragg diffraction in crystals to concentrate
incident photons onto a distant detector. The Laue lens and the detector
are carried by two separate satellites flying in
formation. Significant effort is being devoted to studying different
types of crystals that may be suitable for focusing gamma rays in two
100~keV wide energy bands centered on two lines which constitute the prime
astrophysical interest of the MAX mission: the 511~keV positron
annihilation line, and the broadened 847~keV line from the decay of $^{56}$Co
copiously produced in Type~Ia supernovae. However, to optimize the
performance of MAX, it is also necessary to optimize the detector used
to collect the source photons concentrated by the lens. We address
this need by applying proven Monte Carlo and event reconstruction
packages to predict the performance of MAX for three different Ge
detector concepts: a standard coaxial detector, a stack of segmented
detectors, and a Compton camera consisting of a stack of strip
detectors. Each of these exhibits distinct advantages and
disadvantages regarding fundamental instrumental characteristics such
as detection efficiency or background rejection, which ultimately
determine achievable sensitivities. We conclude that the Compton
camera is the most promising detector for MAX in particular, and for
Laue lens gamma-ray telecopes in general.
}

\maketitle

%
%________________________________________________________________

\section{\label{introduction} Introduction}

Gamma-ray lines carry unique information of prime importance for our
understanding of fundamental astrophysical questions such as the
origin of heavy elements or the mechanisms behind the spectacular
death of stars in supernovae \citep[see e.g.\ reviews
by][]{Prantzos05, Weidenspointner06}. Despite their paramount
interest, observations of gamma-ray lines have historically been plagued
by intense and complex instrumental backgrounds against which the much
smaller signals from celestial sources need to be discerned
\citep[e.g.][]{Weidenspointner05}. Recently, a novel experimental 
technique, promising to improve the sensitivity of existing gamma-ray
telescopes by one to two orders of magnitude, has been demonstrated: the
Laue lens \citep{Halloin04, vonBallmoos04}. 

%Considered physically impossible for a long time {\bf ???? TBD:
%references ???}, 
A Laue lens takes advantage of Bragg diffraction in
crystals to concentrate incident gamma rays onto a detector
\citep{Lund92, Smither95, Halloin04, vonBallmoos04}. In this approach
it is possible to employ a large photon collecting area together with
a small detector volume, which results in a greatly increased
signal-to-background ratio and hence a greatly improved sensitivity.

MAX is a Laue lens telescope that has been proposed to the French
Space Agency CNES in response to an announcement of opportunity for a
formation flight demonstration mission \citep{Barriere06,
vonBallmoos06}. The MAX gamma-ray Laue lens consists of Ge and Cu
crystals and has a focal length of about 86~m. MAX is designed to
concentrate gamma rays in two 100~keV wide energy bands centered on
the two lines which constitute the prime astrophysical interest of the
mission: the 511~keV positron annihilation line, and the broadened
847~keV line from the decay of $^{56}$Co copiously produced in Type~Ia
supernovae.

Significant effort is being devoted to studying different types of
crystals that may be suitable for focusing gamma rays at nuclear line
energies \citep[e.g.][]{Abrosimov06, Courtois06, Smither06}. However,
to achieve the best possible performance of MAX, it is also necessary
to optimize the detector used to collect the source photons
concentrated by the lens. We address this need by applying proven
Monte Carlo and event reconstruction packages to predict the
performance of MAX for three different Ge detector concepts: a
standard co-axial detector, a stack of segmented detectors, and a
Compton camera consisting of a stack of strip detectors. We chose Ge
as detector material since it provides the best energy resolution for
line spectroscopy in the energy range of nuclear transitions. Each of
these detector concepts exhibits distinct advantages and disadvantages
regarding fundamental instrumental characteristics such as detection
efficiency or background rejection, which ultimately determine
achievable sensitivities. Our goal is to identify the most promising
detector concept for a Laue lens.  We consider the expected
sensitivity to be the most important performance parameter, but also
include capabilities for spectroscopy, imaging, and polarimetry in our
final decision. The most promising detector concept will be studied in
more detail and optimized in the future. First advances in the design
of a Compton detector are presented in a companion paper by
\citet{Wunderer06b}.

%\begin{figure*}
%%\centerline{\epsfig{figure=aitoff_iter30.ps,width=13.5cm}}
%\centerline{\epsfig{figure=/users-data/weiden/SPI/PosCont/obs/All0019-0130-pub041210+vela/images/mrem2.ellipse_1.25_-0.75_10_6.CrabCygX1.ds1-f0.od.d.d.box5-5.bulge/aitoff_iter8.ps,bbllx=57pt,bblly=369pt,bburx=508pt,bbury=593pt,clip=,width=15.cm}}
%\caption{A Richardson-Lucy sky map of extended emission in the
%combined Ps analysis intervals 410--430, 447-465, and 490--500~keV. 
%%The emission appears to be symmetric about the Galactic center, its
%%centroid coincides well with the Galactic center. 
%The contour levels indicate intensity levels of $10^{-2}$, $10^{-3}$,
%and $10^{-4}$ ph~cm$^{-2}$~s$^{-1}$~sr$^{-1}$. Details are given in
%the text.}
%\label{fig1}
%\end{figure*}

%________________________________________________________________

\section{\label{sim_performance} Simulation of instrument performance}

This section provides a brief overview of the simulation and analysis
techniques that we employed to estimate by {\it ab
initio} Monte Carlo simulation the performance of three different
detector concepts for MAX. Our study benefited from experience gained
from the modelling of the performance of past or existing gamma-ray
missions such as TGRS on board {\sl Wind}, the Ramaty High Energy Solar
Spectroscopic Imager ({\sl RHESSI}), or SPI onboard {\sl INTEGRAL} by
Monte Carlo simulation \citep[see][ respectively]{Weidenspointner05,
Wunderer04, Weidenspointner03}, and the enhanced set of Monte Carlo and data
analysis tools developed for predicting the performance of various
instrumental concepts for a future {\it Advanced Compton Telescope}
\citep{Wunderer06a,Boggs06a}.

\subsection{\label{sim_performance_instr-sc-models} Instrument and
spacecraft models}

Among other inputs, the simulation of the performance of a gamma-ray
instrument requires a detailed computer description of the
experimental set-up under study. This so-called mass model specifies
the geometrical structure of instrument and spacecraft, the atomic
and/or isotopic composition of materials, and sets parameters that
influence the transport of particles in different materials. 

\begin{figure}[t]
%\fbox{
%\centerline{
\begin{center}
%\fbox{
%\epsfig{figure=/users/weiden/MAX/MassModel/MAX-TGRS/vac0_hiddenlines.ps,bbllx=149pt,bblly=177pt,bburx=411pt,bbury=651pt,clip=,width=6cm}
\epsfig{figure=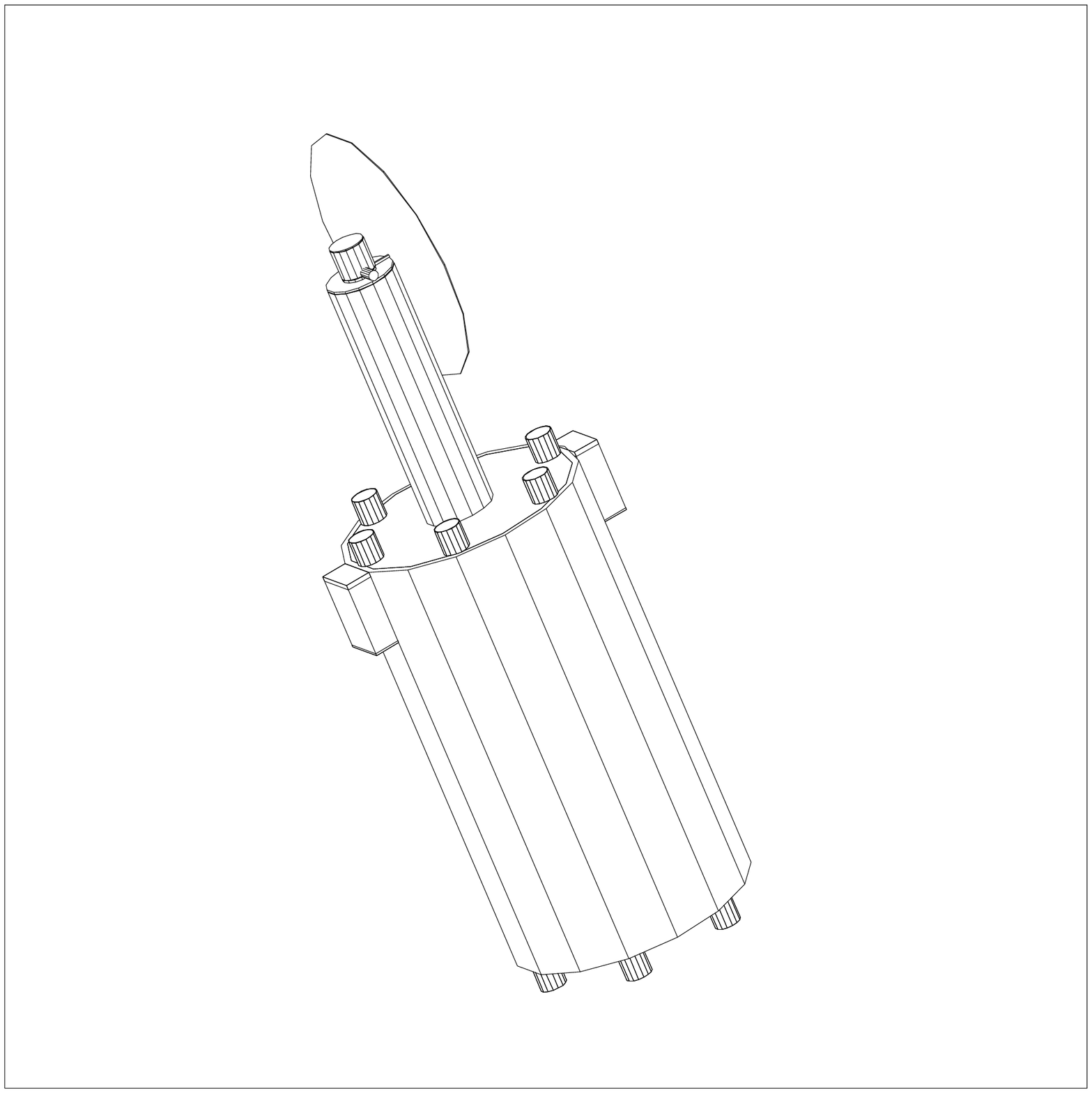,bbllx=149pt,bblly=177pt,bburx=411pt,bbury=651pt,clip=,width=6cm}
%}
\end{center}
%}
%}
\caption{A side view of the MAX detector spacecraft model. The
spacecraft body is cylindrical. The detector is
situated on top of a 1~m tower. The circular radiator assumed to
passively cool the detector is clearly visible. Details are given in
the text.}
\label{overview_fig}
\end{figure}

\begin{figure}[t]
%\fbox{
%\centerline{
\begin{center}
\epsfig{figure=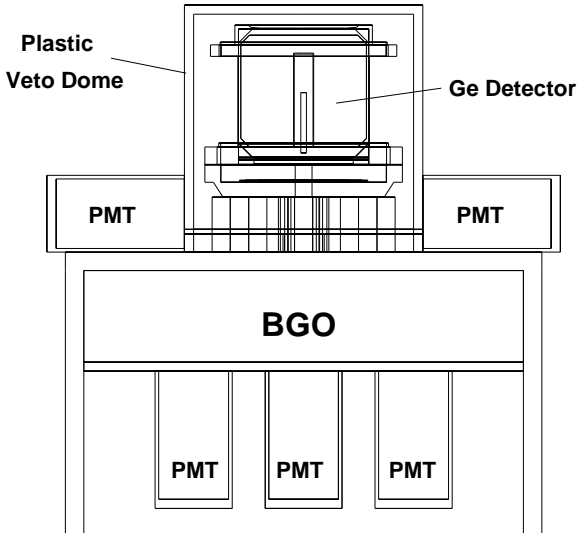,bbllx=122pt,bblly=244pt,bburx=484pt,bbury=570pt,clip=,width=8cm}
\end{center}
%}
%}
%\centerline{\epsfig{figure=/users/weiden/MAX/MassModel/MAX-TGRS/max-tgrs_det-tower-section.ps,width=8cm}}
\caption{A section of the MAX-TGRS detector mass
model, including the top of the tower. The Ge crystal is surrounded by
a plastic veto dome at the sides and at the top, and by a BGO crystal
at the bottom.}
\label{maxtgrs_fig}
\end{figure}

\begin{figure}[t]
%\fbox{
\begin{center}
\epsfig{figure=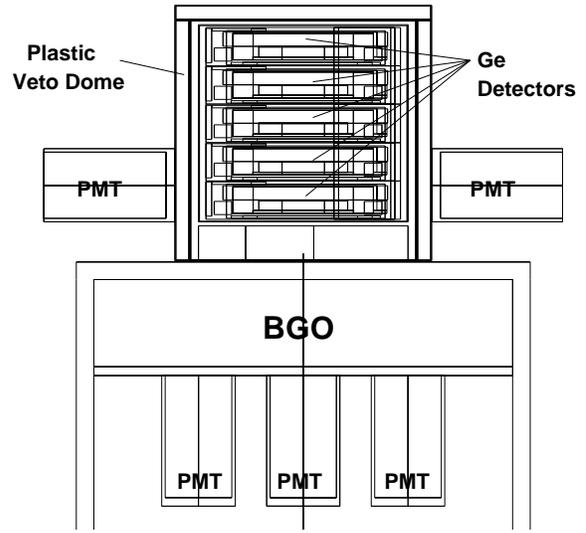,bbllx=110pt,bblly=230pt,bburx=490pt,bbury=570pt,clip=,width=8cm}
\end{center}
%}
\caption{A section of the MAX-NCTseg and MAX-NCTcompt detector mass
models, including the top of the tower. The stack of Ge detectors is
surrounded by a plastic veto dome at the sides and at the top, and by
a BGO crystal at the bottom.}
\label{maxnct_fig}
\end{figure}

The basic design of the MAX detector spacecraft and the mounting and
cooling of the detector concepts assumed in this study emerged from
the CNES pre phase A study of the MAX mission \citep{Barriere06,
vonBallmoos06}.  This basic design is identical for all three detector
concepts, the only difference between the mass models is in the
definition of the detectors. As can be seen in
Fig.~\ref{overview_fig}, in each concept the detector is located on
top of a 1~m tower; passive cooling is provided by a Be radiator of
diameter 1~m.
%; this separation of
%the detector from the spacecraft body is intended to reduce possible
%instrumental background created in the satellite structure. 
The dimensions of the cylindrical spacecraft body are radius 60~cm and
height 192~cm. The spacecraft contains, among other components, tanks
for hydrazine propellant and for cold gas, various electronics boxes,
reaction wheels, and thrusters. The total mass of the spacecraft is
about 260~kg. The tower separating the detector from the spacecraft
body is intended to reduce possible instrumental background created in
the satellite structure. In addition, a 5~cm thick BGO (bismuth
germanate) crystal at the top of the tower and underneath the detector
(see Figs.~\ref{maxtgrs_fig} and \ref{maxnct_fig}) serves as active
(veto) and passive shield for the detector. The combined mass of the
tower structure, the BGO veto shield, the radiator, and various
detector electronics components is about 47~kg.

To study the performance of a standard co-axial detector concept for
MAX (hereafter: MAX-TGRS) we resorted to the TGRS Ge detector flown on
the {\sl Wind} mission. The TGRS detector has been described by
\citet{Owens95}; our mass model is a modified version of the TGRS mass
model used by \citet{Weidenspointner05} for their detailed
instrumental background study. A section of the MAX-TGRS detector mass
model, including the top of the tower, is depicted in
Fig.~\ref{maxtgrs_fig}. Size, geometry, and material composition of Ge
crystal, cathode, and Al housing remained unchanged. For MAX-TGRS, the
radiative cooler of the original TGRS detector was removed. Instead, a
cold finger leading to the radiator was introduced, and miscellaneous
passive materials representing assumed detector support structure and
electronics were positioned below the detector housing.  The detector
assembly is enclosed on the sides and at the top by a 0.5~cm thick
plastic veto shield, which is viewed by two photomultipliers
(PMTs). Also depicted in Fig.~\ref{maxtgrs_fig} are the BGO veto
shield beneath the detector assembly and the respective PMTs. Plastic
dome and BGO crystal cover all lines of sight to the Ge crystal. The
volume of the Ge crystal (height about 6.1~cm, radius about 3.4~cm) is
about 216~cm$^3$, the total mass of the detector assembly including
the plastic veto dome is about 3~kg.

The performance of a stack of segmented detectors and of a Compton
camera consisting of a stack of strip detectors was studied with the
exact same mass model, but different analysis procedures for the
simulated data (see Sec.~\ref{sim_performance_data_analysis}). Both
the segmented (hereafter: MAX-NCTseg) and the Compton detector
(hereafter: MAX-NCTcompt) concepts consist of a stack of five detector
modules modelled after the successfully tested Ge detectors of the
balloon borne {\it Nuclear Compton Telescope} \citep[NCT][]{Boggs06b}.
%In case of
%MAX-NCTseg, we assumed that each detector layer consists of two
%segments: a circular, central segment, and a second segment comprising
%the remaining detector surface. The radius of the central pixel was
%varied in the data analysis . For MAX-NCTcompt we assumed
%
As can be seen in Fig.~\ref{maxnct_fig}, the basic layout of the
instrument geometry (or mass model) for these two concepts is the same
as that of MAX-TGRS: the detector assembly is located inside a plastic veto
dome, with the BGO veto shield below. Each of the five detector planes
is roughly $8 \times 8$~cm$^2$ in size, with a thickness of 1.5~cm,
yielding a total detector volume of about 480~cm$^3$. The gap between
adjacent detector planes was chosen to be 0.7~cm in our concepts. The
total mass of the detector assembly including the plastic veto dome is
about 6~kg.

Each of these detector concepts exhibits distinct advantages and
disadvantages. From a technical point of view, the MAX-TGRS concept is
simplest and easiest to realize, while MAX-NCTcompt is the most
complex and demanding. MAX-TGRS has only one detector channel,
MAX-NCTseg a few, MAX-NCTcompt a few hundred; consequently cooling
MAX-NCTcompt is much more challenging than MAX-TGRS. MAX-NCTcompt
offers superior background rejection capabilities, at
the price of reduced photopeak efficiency, because much more
information is available for each registered event than for the other
two concepts.
%. 
%On the other hand, the photopeak efficiency for MAX-NCTcompt
%is lowest. 

Finally, MAX-NCTcompt has the unique advantage of fine
spatial resolution, which is indispensable for realizing imaging and
polarimetry.

We would like to emphasize that all three detector concepts are
conservative in the sense that we only used detector designs that have
already been flown and successfully operated in a space
environment. 
%Our assumed detectors could therefore be built today and that
%could be built today. 
However, the desgin of all three concepts can be improved, e.g.\ by
minimizing the amount of passive materials, by carefully selecting the
passive materials (e.g.\ elemental composition: carbon fiber instead
of Al structure), or by optimizing the geometry and amount of detector
material for the photon energies of interest to MAX. This is
particularly pertinent for the NCT detectors, which currently are
designed with emphasis on cost as well as reliability and robustness
for use in a balloon demonstration flight. We therefore expect our
performance estimates to be conservative, and that improvements of the
detector designs will result in improved performance.

\subsection{\label{sim_performance_lens-model} Lens beam and
effective area}

\begin{figure}[t]
%\fbox{
%\centerline{\epsfig{figure=MAXbeam.ps,width=7cm}}
\begin{center}
%\fbox{
%\hspace*{-2ex}
%\epsfig{figure=MAXbeam.ps,bbllx=42pt,bblly=108pt,bburx=572pt,bbury=662pt,clip=,width=8cm}
\epsfig{figure=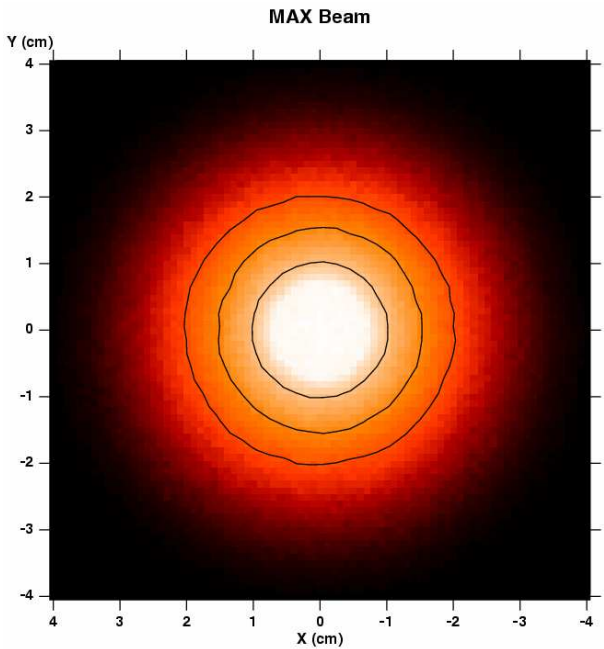,width=8cm}
%}
\end{center}
%}
\caption{The focal spot distribution of photons concentrated by the
MAX Laue lens onto the detector plane. The three contour levels
indicate, with increasing radius, the detector surface exposed to
50\%, 75\%, and 90\% of all incident photons, respectively. More
details are given in the text.}
\label{maxbeam_fig}
\end{figure}

%In addition to models of the experimental set-up, 
Estimating the performance of MAX detector concepts also requires a
model for the focal spot distribution of photons concentrated by the
Laue lens, and an estimate of its effective area (i.e.\ the
geometrical area times the diffraction efficiency).
%
%The focal spot distribution assumed in this study has been 
For this study, both quantities were determined
by Monte Carlo simulation.
% \citep[see][ for more detail]{Barriere06}.
% and is depicted in Fig.~\ref{maxbeam_fig} \citep{Barriere06}. 
As described in more detail in \citet{Barriere06}, for this study the
lens crystals were assumed to have a geometrical size of 1.5~cm
$\times$ 1.5~cm and a mosaicity of 30$^{\prime\prime}$. The focal
length was assumed to be 86~m, and the source was assumed to be
located on the optical axis (i.e.\ the lens is pointed at the source).
%{\bf ??? TBD: references ???}. {\bf ????
%which crystal planes ???}. 

For these parameters, about 50\% of the photons from an on-axis point
source that are diffracted by the lens are concentrated within 1~cm
from the center of the focal spot, as can be seen in
Fig.~\ref{maxbeam_fig} depicting the simulated focal spot
distribution. To actually perform the detector response simulations in
our study, the Laue lens focal spot distribution was introduced as a
new beam type named {\tt LRAD} into the MGGPOD Monte Carlo suite
described below in Sec.~\ref{sim_performance_mggpod}.
%The input required for the response
%simulations is the number of photons per annulus for a user-defined
%radial binning.
%
Unlike the focal spot distribution of the Laue lens design considered
here,
% \citep{Barriere06}, 
the effective area of the lens is a function
of energy: about 1191~cm$^2$ and 661~cm$^2$ at energies of 511~keV and
847~keV, respectively.

\subsection{\label{sim_performance_radenv-models} Radiation environment
models}

In the CNES pre phase A study it was concluded that MAX would be best
operated in either a high Earth orbit (HEO) or an L2 orbit
\citep{vonBallmoos06}. The instrumental background would then mainly
be due to two radiation fields, namely diffuse cosmic gamma rays and
Galactic cosmic rays. Both radiation fields were assumed to be
isotropic in our simulations. The spectrum of the diffuse cosmic
gamma-ray background was taken from the analytical description given
by \citet{Gruber99}. The spectrum and intensity of Galactic cosmic-ray
protons was modelled using the {\tt COSN} default solar minimum
spectrum of the MGGPOD package, which is based on the cosmic-ray
propagation models of
\citet{Moskalenko02}.

Galactic cosmic rays not only produce prompt background due to
hadronic interactions and de-excitations of excited nuclei, but also
produce radioactive isotopes whose decay gives rise to delayed
instrumental background. When simulating this delayed background, we
assumed that the instrument and spacecraft materials had been
irradiated for one year with the {\tt COSN} cosmic-ray proton spectrum
and intensity.

\subsection{\label{sim_performance_mggpod} MGGPOD Monte Carlo suite}

We used the MGGPOD Monte Carlo package \citep{Weidenspointner05} to
simulate the response and the instrumental background expected for
each of the three different detector concepts for
MAX. MGGPOD is a user-friendly suite of Monte Carlo codes that is
available to the public from a site at
CESR\footnote{http://sigma-2.cesr.fr/spi/MGGPOD/}. MGGPOD is built
around the widely used GEANT3.21 package \citep{Brun95} and allows 
simulation {\it ab initio} of the physical processes relevant for
estimating the performance of gamma-ray instruments. Of particular
importance is the production of instrumental backgrounds, which
include the build-up and delayed decay of radioactive isotopes as well
as the prompt de-excitation of excited nuclei, both of which give rise
to a plethora of instrumental gamma-ray background lines in addition
to continuum backgrounds. Among other packages, MGGPOD includes the
GLECS \citep{Kippen04} and GLEPS \citep{McConnell_Kippen04} packages for
simulating the effects of atomic electron binding and photon
polarization for Rayleigh and Compton scattering.

As mentioned in Sec.~\ref{sim_performance_lens-model}, for this study
a new beam type named {\tt LRAD} was introduced into the MGGPOD Monte
Carlo suite. This beam allows the user to define an azimuthally
symmetric incident photon flux which is characterized by its radial
profile. The direction of incidence is assumed to be identical for all
photons, rather than spread over the directions to the lens covering a
few degrees in the field-of-view, which is an approximation that
should not signficantly affect our detector performance estimates at
this early stage of the study.

%The input required for the response simulations is the number of
%photons per annulus for a user-defined radial binning.

\subsection{\label{sim_performance_megalib} MEGAlib analysis package}

The complex event analysis for the MAX-NCTcompt detector concept was
performed with the MEGAlib package \citep{Zoglauer06}. Originally, it
had been developed for the MEGA Compton telescope prototype 
%which consists of a tracker made of thin Si layers and a CsI calorimeter
\citep{Kanbach04}. The package provides the complete data analysis
chain for Compton telescopes, including the crucial steps of event
reconstruction and background rejection, which are described in more
detail in \citet{Zoglauer05, Zoglauer06, Wunderer06b} and references therein.

\subsection{\label{sim_performance_data_analysis} Data analysis}

We compared the performance of the three MAX detector concepts under
study for three different gamma-ray lines: narrow lines at 511~keV and
847~keV, and a broadened line (3\% full width at half maximum, FWHM, deemed
typical for Type~Ia supernovae) at 847~keV. For each concept we
simulated the instrumental response to these three lines for an
on-axis point source. We also simulated the instrumental backgrounds
due to diffuse cosmic gamma rays, to Galactic cosmic-ray protons at solar
minimum, and to the decay of radioactive isotopes resulting from
one year of cosmic-ray proton irradiation. For all three concepts
radioactive decays in the detectors and diffuse cosmic gamma rays were
found to be the dominant instrumental background components. In
comparison, the prompt cosmic-ray induced background is small, and the
background due to radioactive decays in the satellite structure is
even smaller.

Despite the fact that source photons are concentrated by the Laue lens
onto the detector, MAX is still largely background dominated (the
signal-to-noise ratio being on the order of several per cent), and we
calculated its sensitivity to an on-axis gamma-ray line point source
according to
\begin{equation}
\label{sens_bgddom_unknown_final}
f_{n_\sigma} = \frac{n_\sigma \cdot \sqrt{\sum_{i=1}^{n_b} b_i(\Delta
  E)}} {A_{\rm eff}  \cdot \epsilon(\Delta E)  \cdot \sqrt{t_{tot}}} 
\cdot \eta
\end{equation}
where $f_{n_\sigma}$ is the sensitivity in [ph~cm$^{-2}$~s$^{-1}$],
$n_\sigma$ is the statistical significance of the detection,
$\sum_{i=1}^{n_b} b_i(\Delta E)$ is the sum of all instrumental
background components in [cts~s$^{-1}$] in the analysis interval
$\Delta E$ centered on the line energy, $A_{\rm eff}$ is the effective
area of the Laue lens in [cm$^2$],
% (i.e.\ the geometrical area times the diffration efficiency at the
%line energy), 
$\epsilon(\Delta E)$ is the photopeak efficiency, $t_{tot}$ is the
total effective observation time in [s], and $\eta$ is a factor in the
range 1--2 whose value depends on how the instrumental background
during an observation is determined. Ideally, the instrumental
background is known, and $\eta$ becomes 1. If the instrumental
background is determined through an on-off observation strategy, i.e.\
one half of the total effective observation time is spent pointing at
the source, and the other half pointing away from the source measuring
the instrumental background, $\eta$ is 2. An intermediate case can be
realized by operating two detectors simultaneously such that they
alternately point at the source and away from it; $\eta$ then assumes
a value of $\sqrt{2}$.

For MAX-TGRS data analysis is straight forward since the only event
selections that can be applied in the case of a single detector
crystal are the thresholds of the detector and of the veto shields and
the width of the analysis energy interval. We assumed the same
thresholds for all three concepts: 15~keV for the detector, and veto
thresholds of 70~keV and 200~keV for the BGO and plastic dome shields,
respectively. We assumed an energy resolution as measured for the SPI
detectors \citep{Lonjou05} for the MAX-TGRS detector. 
In the MAX-TGRS concept it is impossible to separate source signal and
instrumental background from a single observation; an on-off pointing
strategy must be adopted, and $\eta = 2$ in
Eq.~\ref{sens_bgddom_unknown_final} when calculating the instrument
sensitivity. At best, two MAX-TGRS detectors could be operated
simultaneously; the minimum value of $\eta$ therefore is $\sqrt{2}$
for this concept.

\begin{table*}[t]
%\caption[]{}
\caption{The sensitivity of three detector concepts for MAX for three
different gamma-ray lines. Sensitivities are for a statistical
significance of $3\sigma$ and a total effective observation time of
$10^6$~s. The effective area of the MAX Laue lens was assumed to be
1191~cm$^2$ and 661~cm$^2$ at 511~keV and 847~keV, respectively.
The quoted values pertain to the energy interval $\Delta E$,
centered on the line energy, that optimizes the sensitivity. The ranges
in sensitivity reflect the possible values of $\eta$ in
Eq.~\ref{sens_bgddom_unknown_final} as discussed in the text. 
%for MAX-TGRS reflects the difference between the cases
%of unknown (see Eq.~\ref{sens_bgddom_unknown_final}) and known (see
%Eq.~\ref{sens_bgddom_unknown_final2}) instrumental background. 
The range in sensitivity for MAX-NCTseg in addition includes the two
choices for treating energy deposits in unused detectors, which may be
ignored or used as additional veto. 
%The sensitivity of MAX-NCTcompt
%was calculated assuming that the instrumental background is known,
%i.e.\ can be determined from the source observation (see
%Eq.~\ref{sens_bgddom_unknown_final2}).
}
\label{table_sens}
\begin{center}
\begin{tabular}[t]{lccc}
\hline 
\noalign{\smallskip}
 & MAX-TGRS & MAX-NCTseg & MAX-NCTcompt \\ 
\noalign{\smallskip}
\hline \hline
\noalign{\smallskip}
Line Energy [keV] &  \multicolumn{3}{c}{Sensitivity
[$10^{-6}$~ph/cm$^2$/s]} \\  
\noalign{\smallskip}
\hline \hline
\noalign{\smallskip}
%511 & $(3.0-6.0) \times 10^{-6}$ & $(1.8-4.5) \times 10^{-6}$ & $1.3 \times
%10^{-6}$ \\
%847 & $(3.5-6.9) \times 10^{-6}$ & $(1.9-3.6) \times 10^{-6}$ & $1.3 \times
%10^{-6}$ \\
%847 (3\% FWHM) & $(1.3-2.5) \times 10^{-5}$ & $(0.7-1.5) \times
%10^{-5}$ & $3.5 \times 10^{-6}$ \\
511 & 4.2--6.0 & 2.5--4.6 & 1.3--1.8 \\
847 & 4.9--6.9 & 2.6--3.8 & 1.3--1.8 \\
847 (3\% FWHM) & 18--25 & 10--15 & 3.5--4.9 \\
\noalign{\smallskip}
\hline
%\noalign{\smallskip}
\end{tabular}
\end{center}
\end{table*}

\begin{table*}[t]
%\caption[]{}
\caption{The photopeak efficiencies and the background rates that went
into the calculation of the line sensitivities in
Table~\ref{table_sens}.}
\label{table_eff_bgd}
\begin{center}
\begin{tabular}[t]{lccc}
\hline 
\noalign{\smallskip}
 & MAX-TGRS & MAX-NCTseg & MAX-NCTcompt \\ 
\noalign{\smallskip}
\hline \hline
\noalign{\smallskip}
Line Energy [keV] &  \multicolumn{3}{c}{Photopeak Efficiency [\%]} \\ 
\noalign{\smallskip}
\hline \hline
\noalign{\smallskip}
511 & 38 & $22-28$ & 6 \\
847 & 24 & $16-22$ & 6 \\
847 (3\% FWHM) & 27 & $17-24$ & 6 \\
\noalign{\smallskip}
\hline \hline
\noalign{\smallskip}
Line Energy [keV] &  \multicolumn{3}{c}{Background Rate [cts/s]} \\ 
\noalign{\smallskip}
\hline \hline
\noalign{\smallskip}
511 & $2.1 \times 10^{-1}$ & $2.3\times10^{-2} -
6.5\times10^{-2}$ & $1.0\times10^{-3}$ \\
847 & $3.4 \times 10^{-2}$ & $4.2\times10^{-3} -
8.6\times10^{-3}$ & $2.6\times10^{-4}$ \\
847 (3\% FWHM) & $5.2 \times 10^{-1}$ & $8.0\times10^{-2} -
1.6\times10^{-1}$ & $2.1\times10^{-3}$ \\
\noalign{\smallskip}
\hline
%\noalign{\smallskip}
\end{tabular}
\end{center}
\end{table*}

For MAX-NCTseg, data analysis is slightly more complicated. As
described in Sec.~\ref{sim_performance_lens-model}, for an on-axis
point source the Laue lens concentrates source photons onto a
relatively small focal spot. This can be exploited in the data
analysis by including the criterion that a valid event must
deposit energy in a cylindrical detector volume 
%of adjustable radius
centered on the optical axis in a selected set of detector layers;
events that do not deposit energy in these central detector volumes
are most likely instrumental background that should be rejected. We
implemented this simple scheme by assuming that each detector layer
consists of two segments or pixels: a cylindrical, central segment,
and a second segment comprising the remaining detector layer
volume. 
%The radius of the central pixel was varied in the data analysis. 
Different values for the central radius were tried.
In addition, we varied in the analysis the number of detector layers
used to record source photons (source recording detector layers --
SRDLs) in order to estimate the optimum number of detector layers for
the MAX-NCTseg concept without performing a full simulation for each
possibility. In doing so, we also had to choose how to treat remaining
detector layers (background recording detector layers -- BRDLs):
these were either ignored or used as additional veto shields. A valid
event was required to deposit more than 15~keV in at least one central
pixel of the SRDLs
%used to record source photons 
without any veto trigger. We assumed an energy resolution as
measured for the NCT detectors (S.\ Boggs, priv.\ comm.). For the same
reasons given for the MAX-TGRS concept, $\eta$ lies in the range
$\sqrt{2}$--2 for the MAX-NCTseg concept.
%As for MAX-TGRS, the sensivity of MAX-NCTseg was determined
%using Eq.~\ref{sens_bgddom_unknown_final}.

Data analysis for MAX-NCTcompt was most complex and performed using
MEGAlib. Many logical criteria can be applied to decide whether the
energy deposits in the detector are consistent with an interaction
sequence of a photon originating from the Laue lens, including the
criterion that the first interaction needs to occur in a cylindrical
detector volume of a given radius centered on the optical axis in any
one of a selected set of detector layers
\citep[see][ for details]{Wunderer06b}.
%Again, we have a choice for the number of detector layers
%used to record source photons, but in addition many logical criteria
%can applied to decide whether the energy deposits in the detector are
%consistent with an interaction sequence of a photon originatig from
%the Laue lens, including the criterion that the first interaction
%needs to occur in a circular region of adjustable radius centered on
%the optical axis \citep[see][ for details]{Wunderer06b}. 
The spatial pitch of the Ge strip detectors was assumed to be 2~mm in
the plane of the detectors, and 0.4~mm in depth (S.\ Boggs, priv.\
comm.). Again, we assumed an energy resolution as measured for the
NCT detectors. 
The inherent imaging capabilities of the MAX-NCTcompt detector should
permit both the source signal and the instrumental background to be
measured in a single observation, as was the case for the imaging
Compton telescope COMPTEL \citep{Schoenfelder93}. Any need for
off-source observations is therefore obviated, and the total effective
observation time $t_{tot}$ can be spent pointing at the source. In
this case the value of $\eta$ approximates 1 in
Eq.~\ref{sens_bgddom_unknown_final}. An exact determination of the
value of $\eta$ is difficult and beyond the scope of this paper. We
expect the number of data space bins free of source signal to exceed
that of data space bins containing source signal, hence $\eta$ should
be smaller than $\sqrt{2}$. We
%therefore 
conservatively adopt a range of
1--$\sqrt{2}$ for $\eta$ for the Compton detector concept.

%Then the sensitivity
%is given by
%\begin{equation}
%\label{sens_bgddom_unknown_final2}
%f_{n_\sigma} = \frac{n_\sigma \cdot \sqrt{\sum_{i=1}^{n_b} b_i(\Delta
%  E)}} {A_{\rm eff} \,   \epsilon(\Delta E) \,   \sqrt{t_{tot}}}
%\end{equation}
%with all of the observation time $t_{tot}$ being spent on target.

%________________________________________________________________

\section{\label{results} Results}

The sensitivities of the three different MAX detector concepts for an
on-axis point source for three different gamma-ray lines are
summarized in Table~\ref{table_sens}. Sensitivities are for a
statistical significance of $3\sigma$ and a total effective
observation time of $10^6$~s; the ranges reflect the possible values
of $\eta$ in Eq.~\ref{sens_bgddom_unknown_final} as discussed
above. The effective area of the MAX Laue lens was assumed to be
1191~cm$^2$ and 661~cm$^2$ at 511~keV and 847~keV,
respectively. Sensitivity values are quoted for the best choices of
both the radius of the central pixel and of the width of the energy
interval $\Delta E$ centered on the line energy.
%{\bf ??? event selections, detector layers used, and energy interval
%$\Delta E$. ???}
%The quoted values pertain to the
%radius of the central pixel, and the energy interval $\Delta E$
%centered on the line energy, that optimize the sensitivity.

%The range in sensitivity for MAX-TGRS
%reflects the difference between the cases of unknown (see
%Eq.~\ref{sens_bgddom_unknown_final}) and known (see
%Eq.~\ref{sens_bgddom_unknown_final2}) instrumental background. 

The range in sensitivity for MAX-NCTseg in addition includes the two
choices for treating energy deposits in BRDLs, which
may be ignored or used as additional veto, and are given for the
optimal choice of SRDLs in each case. If energy deposits in
BRDLs are ignored, for the 511~keV line it is best to
use all five detector layers; for the 847~keV lines there is little
difference between using three, four, or five layers (values are
quoted for five layers).  If BRDLs are used as
additional veto the sensitivity can be slightly improved. It is then
best to use only three detector layers in this case for both the
511~keV and the 847~keV lines. In either case the optimal central
pixel radius is about 1.1~cm.

For MAX-NCTcompt the best radial size of the detector region where the
first interaction needs to occur is slighty larger than for
MAX-NCTseg;
%: about 1.2~cm for the 511~keV line, and about 1.5~cm for
%the broadened 847~keV line. 
values range between 1.2 and 1.5~cm, depending on the details of the
event selections. The achieved sensitivity depends only weakly on the
choice of detector layers in which the first interaction needs to
occur. It seems that restricting the first interaction to the top four
layers is best.

As can be seen from Table~\ref{table_sens}, the Compton detector
concept MAX-NCTcompt offers the best sensitivity for each of the three
lines. In order to illustrate fundamental performance characteristics
such as detection efficiency or background rejection,
Table~\ref{table_eff_bgd} summarizes the photopeak efficiencies and
the background count rates corresponding to the choices for energy
band, detector layers, and event selections that optimize the
sensitivity for each of the three detector concepts and all three
lines under study. The ranges for the MAX-NCTseg concept reflect the
two different treatments of energy deposits in BRDLs
(the lower and upper bounds are obtained if BRDLs are
treated as additional veto or ignored, respectively). The photopeak
efficiencies for MAX-NCTseg do not fall far short of those obtained
with MAX-TGRS. Differences are due to the fact that photons can more
easily escape MAX-NCTseg than MAX-TGRS, and that only a fraction of
the incident photons interacts in one of the central pixels of the
segmented MAX-NCTseg detectors. In contrast, the MAX-NCTcompt
photopeak efficiency is much smaller. For this concept the rather low
photopeak efficiency is due to the severe event selections, which
result in many source events being rejected.  Nevertheless, the
MAX-NCTcompt concept offers the best sensitivity because of its
superior capabilities for rejecting instrumental background, as can be
seen in Table~\ref{table_eff_bgd}.

%________________________________________________________________

\section{\label{summary} Summary and conclusion}

We have used {\it ab initio} Monte Carlo simulations to compare the
performance of three different Ge detector concepts for the MAX Laue lens
gamma-ray telecope: a standard co-axial detector, a stack of segmented
detectors, and a Compton camera consisting of a stack of strip
detectors. The performance was assessed for an on-axis point source in
three different gamma-ray lines: narrow lines at 511~keV and 847~keV,
and a broadened (3\% FWHM) line at 847~keV.

We find that the Compton detector concept MAX-NCTcompt offers the best
sensitivity for each of the three lines. The Compton concept also
offers other unique advantages over the other two concepts. Because of
their fine spatial resolution, the detectors of a Compton camera are
%in addition 
ideally suited to follow the inevitable small excursions of the focal
spot on the detector surface due to residual relative motions of the
lens and detector spacecraft; with a Compton camera one could also
adjust the size of the focal spot to the requirements of a given
observation during data analysis. The fine spatial resolution
necessary for Compton detectors is also required if the limited
imaging capabilites of a Laue lens are to be exploited, e.g.\ to
separate close point sources or to study the morphology of slightly
extended emission such as that from Galactic supernova remnants.
%
%Finally, a
%Compton detector is intrinsically perfectly suited to perform
%polarimetry {\bf ??? ??? REFERENCE ??? ???}, which would open a new
%observational window on many gamma-ray sources in which strong
%magnetic fields are present, such as pulsars, or on jets expelled by
%compact, accreting objects. 
Finally, the complementary characteristics of a Laue lens and of a
Compton detector with respect to photon polarisation render their
combination a powerful polarimeter. At nuclear line energies a Laue
lens does not change the polarisation of the diffracted photons
\citep{Halloin_Bastie06, Halloin06}, while a Compton detector is
instrinsicaly ideally suited for performing polarimetry because of the
azimuthal variation of the scattering direction for linearly polarized
photons \citep{Lei97}. The combination of a Laue lens with a Compton
detector will thus open a new
observational window on many gamma-ray sources in which strong
magnetic fields are present, such as pulsars, or on jets expelled by
compact, accreting objects.
We therefore conclude that a Compton
camera is the most promising detector concept for MAX. We expect this
conclusion to apply not only to the three gamma-ray lines studied
here, but to all Laue lens gamma-ray telescopes proposed for the
nuclear line region, such as the Gamma-Ray Imager
\citep[GRI,][]{Knoedlseder06}.
%the roughly 0.1--10~MeV energy
%range of nuclear transitions. We therefore expect a Compton camera to
%be the best choice for all Laue lens gamma-ray telescopes in the
%nuclear line region.

Although not the primary focus of this study, it is still worth
pointing out that even with a rather conservative design of the
Compton camera that leaves still ample room for improvement, narrow
line sensitivities of about $10^{-6}$~ph~cm$^{-2}$~s$^{-1}$ are
possible with a relatively small mission such as MAX -- an improvement
over the currently best gamma-ray spectrometer SPI onboard the
INTEGRAL observatory of more than an order of magnitude.

There are many aspects in which the Compton camera studied here can be
improved. First steps towards 
%improving 
optimizing the design of the MAX-NCTcompt
detector are presented in a companion paper by
\citet{Wunderer06b} (there, MAX-NCTcompt is referred to as the SMALL
design). Possible improvements include 
%an improved 
a revised design of the BGO
veto shield to decrease the instrumental background contribution of
cosmic diffuse photons, the reduction of passive materials around the
Ge wafers (passive material is a source of instrumental background and
in addition reduces the photopeak efficiency because some fraction of
the source photon's energy might be deposited there), the careful
selection of passive materials used (e.g.\ elemental
composition), the optimization of the geometry and the spatial and
spectral resolution of the Ge detectors to increase the photopeak
efficiency, or improvements of event reconstruction
algorithms. Efforts to optimize the performance of a Compton detector
for Laue gamma-ray lenses are ongoing \citep[see e.g.\ the companion
paper by][]{Wunderer06b} and will be reported in future
publications.

%------------------------------------------------------------------

%
%%%% bibliography %%%%%%%%%%%%%%%%%%%%%%%%%%%%%%%%%%%%
%

%\vspace*{-3ex}

\end{document}